# Accounting for Trust and Safety: Bounding Content-Moderation Quality from Records You Already Hold

Aayush Agarwal, Daron Veer Singh Sohi, and Linda Cornelison

Meta

*Author note: This is theoretical work only, and no evaluation was conducted.*

## Abstract

Platforms report content-moderation quality, typically precision, from an expert audit of a random sample of enforcement decisions. That figure is produced inside the system it measures and is rarely checked against anything. We show that a platform's own operational records place a one-sided constraint on it that requires no sample. Successful appeals are a lower bound on wrongful removals: each is a removal which the platform itself reversed on the original evidence. If that count alone exceeds the platform's declared error budget, the realised cohort did not meet the declared target, established from two integers the platform already holds. Confirmed reports give the corresponding result for recall, which holds at any reporting rate. Both bounds carry an explicit parameter for the share of counted concessions that are not genuine, and neither is treated as exact. These are record-based, one-sided consistency checks: they compare the platform's own operational records with its own published target and do not depend on audit sampling. We then show that the natural way to use an audit here does not work. Comparing the reversal count against an audit-derived upper bound for the same quantity yields an alarm whose rejection probability is at most the audit's miscoverage rate, uniformly in the truth; this is the crossing-bound specification test familiar from partial identification, and it tests the measurement rather than the system. Appeal participation and latency affect how often a finding appears and never whether it is true. Second-review error and population mismatch affect validity, and both are inspectable without a behavioural model. A contradiction warrants investigation of population alignment, disposition classification, data pipelines and target provenance; its absence does not validate quality, and we require a pre-specified reach criterion so that a slice the check could never have diagnosed is reported as uninformative rather than clean. We then ask what the transparency regime makes falsifiable. The Digital Services Act compels disclosure of a realised accuracy indicator and of the methodology behind it, but nowhere requires a target declared in advance, and it leaves the term accuracy undefined; where accuracy is defined from complaint outcomes the check becomes tautological, and where no target is declared it has nothing to bind to. We give a conformance profile stating the minimum a published account must contain to be checkable, together with the tradeoffs of requiring it. We report no results; a prospectively specified evaluation protocol is given.

---

# 1. Introduction

Automated systems decide essentially every eligible item within a defined moderation scope unless human review or oversight is required by law. Human effort has moved from deciding cases to *measuring*: sampling decisions, estimating quality, and reporting a figure to leadership and, under the Digital Services Act, to regulators.

That figure is produced by people sharing a rulebook, a training pipeline and tooling with the system they measure. If the rulebook is wrong, or the sampling frame does not represent the population it claims to, the figure can be wrong in ways invisible from inside the measurement process.

Appeals are different in kind, although the second review is not an independent observer: it shares the platform's policy, training and tooling. The decision to contest a removal is made outside the measurement apparatus, and a qualifying reversal is the platform's authoritative concession under the same decision problem, not external ground truth. This paper asks what the benefit of the signal is. It is minimally beneficial as an estimate of error volume, because user feedback is severely selective (Begg and Greenes 1983; Lakkaraju et al. 2017). It is highly beneficial as a bound.

The distinction is the paper's organising idea. An estimate needs the appealing population to resemble the full population. A bound does not. Selection makes a bound loose; it cannot reverse it.

## 1.1 Contributions

1. **Record-based, one-sided bounds** (§§4–5 and §8). Qualifying appeal overturns lower-bound removals inconsistent with the platform's authoritative standard, while qualifying report-driven corrections lower-bound misses. Each side carries its own concession-validity parameter, $\varepsilon$ and $\delta$, so neither direction is assumed exact. These bounds require no model of who appeals or reports and can contradict a target declared in advance for the realised cohort (A5).

2. **A characterisation of validity and diagnostic reach** (§6). Population alignment, decision equivalence and disposition classification determine validity; participation and latency determine how often the bounds become informative.
3. **An applied negative result for the natural audit-based comparison** (§7). A known crossing-bound result shows why the obvious audit comparison tests the measurement and its assumptions rather than moderation quality. The contribution is the diagnosis in this setting, not the underlying inequality.
4. **A checkable reporting specification** (§§11–12). We identify what the Digital Services Act reporting regime lacks for these bounds to be externally checkable and provide a conformance profile and prospectively specified evaluation protocol.

Section 2 places the work in the literature. Sections 3 to 5 develop the bounds and the check. Section 6 separates the real threats from the apparent ones. Section 7 gives the negative result. Sections 8 to 10 cover recall, future evaluation and limitations; §11 asks what the reporting regime makes falsifiable; §12 and §13 give a proposed protocol and conclusions.

# 2. Related work

**Systemic accountability for moderation.** A line of platform-governance scholarship argues that moderation oversight must move from adjudicating individual posts to evaluating systems statistically, and that existing transparency reporting does not support this. This paper is a technical contribution to that agenda: it asks what a reported error rate can be held to.

**DSA reporting consistency.** Sekwenz, Wagner and de Bruijn (2026) develop and apply a three-level framework—internal, external and historical consistency—to Instagram transparency reports, Statement of Reasons data, a systemic risk assessment and an independent audit. Their framework treats discrepancies as signals of reporting anomalies and traceability gaps requiring further inquiry, not as direct measures of moderation error. The present paper asks a narrower question: whether cohort-matched appeal reversals and confirmed report-driven corrections impose one-sided bounds on error counts that can falsify a target fixed in advance. Sekwenz et al. do not derive those bounds, model concession error through $\varepsilon$ or $\delta$, specify diagnostic reach, or analyse the crossing of an operational lower bound with an audit confidence bound.

**Content-moderation accuracy reporting.** Wei, Zufall and Jia (2024) operationalise DSA "accuracy" as precision and recall and address estimating each from random samples; they do not use appeals. Schwemer (2024) observes that complaint counts understate error because not all wrongful decisions are appealed, and proposes working with an estimate of the unappealed share. This paper takes the opposite route: for falsification the estimate is unnecessary, and avoiding it is what makes the result robust. Schwemer is the closest prior work on using complaint outcomes to reason about error volume and does not state the check in §5.

**Partial identification.** The constraints here form an identified set for the wrongful-removal count given a sampled interval and a logical lower bound, which is standard territory (Manski 1989, 2003; Imbens and Manski 2004; Chernozhukov, Lee and Rosen 2013). We claim no novelty in the inferential machinery. In particular, the alarm analysed in §7 — reject when a lower bound crosses an upper bound — is the specification test for partially identified models studied by Bugni, Canay and Shi (2015) as "Test BP", and the standard reading of such a test is the one we reach in §7.2: it interrogates the maintained assumptions, not the parameter. Our contribution there is the applied diagnosis, not the inequality.

**Audit and assurance.** Projecting sampled errors, adding separately known exceptions, and reconciling independent records are standard assurance operations (Horvitz and Thompson 1952; PCAOB AS 2315; ISA 530). The reversal count is analogous to a *known exception* rather than a projected one, which is why it carries no sampling error; §5 is recognisably the assurance operation "known misstatement exceeds tolerable misstatement", transplanted.

**Selective verification.** Begg and Greenes (1983) and Lakkaraju et al. (2017) establish why a selectively verified subset cannot be generalised to a population. We rely on that rather than contradict it: it is why we take a bound instead of an estimate.

**Item-level comparison.** Ames, Handan-Nader, Ho and Marcus (2019) study a natural experiment in which 5–10% of draft opinions at the US Board of Veterans' Appeals were randomly selected for quality review, and compare those determinations against later appellate outcomes on the same cases; their headline finding is that quality review had no appreciable effect on remands. The US Department of Labor's Denied Claims Accuracy programme has maintained equivalent linked records since 2001. We do not propose an item-level join: at moderation sample sizes and reversal rates the overlap is generally too small to support one, which is why population-level bounds are the right instrument here.

**Capture–recapture.** Multiple-systems estimation would use the overlap between audit-detected and appeal-detected errors to estimate the total (Chandra Sekar and Deming 1949; Fienberg 1972). It requires independence and homogeneous capture. Here the mechanisms are positively dependent — an obviously wrong removal is both more likely to be judged wrong on audit and more likely to be appealed — and capture probability varies sharply. The estimate is unstable to the point of uselessness.

# 3. Setting, notation and assumptions

We work within a **declared slice**: one policy area, market, language, product surface, model version, enforcement path, and time window. Results do not generalise across slices; comparing quantities from different slices manufactures false findings.

Throughout, “wrong”, “correct” and “precision” are relative to the platform's frozen authoritative policy standard under the same evidence and decision bar; they do not denote external or normative ground truth.

| Symbol | Meaning | Kind |
| --- | --- | --- |
| $D$ | automated enforcement decisions (removals) in the slice | counted fact |

| Symbol | Meaning | Kind |
| --- | --- | --- |
| D_a | of those, the subset eligible for appeal | counted fact |
| R | removals reversed after a successful appeal | counted fact |
| C | distinct items left up, then removed after a confirmed report under the same evidence, policy and decision bar | counted fact |
| b | the platform's declared error budget, 1 − precision target | declared |
| F | removals inconsistent with the frozen authoritative standard | unobservable |
| p | precision relative to that standard, (D − F)/D | unobservable |
| $\hat{p}$, $[p^-, p^+]$ | audit estimate and confidence interval at level $1 - \alpha$ | judgment, fallible |
| ε | proportion of counted reversals that are not genuine concessions under A2–A3 | unobservable, boundable |
| δ | proportion of counted report-driven corrections that are not genuine misses under the frozen standard | unobservable, boundable |

Counted facts and judgments are kept in separate notation throughout; conflating them causes the error corrected in §8.1.

**Cohort accounting.** D is a cohort of decisions, and R counts only *reversals of decisions in that cohort*, whenever they arrive. Under calendar-window accounting — reversals recorded in a window, regardless of when the underlying decision was made — reversals can attach to decisions outside D and the bounds below fail. This requirement is not optional and is the first thing to verify.

**Deduplication.** R counts at most one reversal per decision. C counts one case per (item × harm type); additional reports of the same item and harm after a final decision create no additional observation.

## 3.1 Assumptions

1. **A0 (containment).** Every counted reversal resolves to a decision within D. Violations are the subject of §6.3.
2. **A1 (injectivity).** Distinct counted reversals map to distinct decisions.
3. **A2 (same decision problem).** The second review saw the same evidence, context and policy version, and applied the same decision bar. New evidence makes it a new decision, out of scope. A deliberately more permissive appeal standard also violates A2: the framework does not apply to a platform that restores when in doubt against an audit that does not.

4. **A3 (adjudication competence).** A genuine concession identifies an index removal as inconsistent with the frozen authoritative standard. §4 relaxes perfect classification by allowing an $\varepsilon$ share of counted reversals not to be genuine concessions.
5. **A4 (coverage).** Where an audit interval is used, it has one-sided frequentist coverage for the realised finite-population quantity, and its sample selection does not depend on appeal status.
6. **A5 (target interpretation).** The declared error budget $b$ applies to the realised finite-population precision $p$ of the declared cohort and is fixed before outcomes in that cohort are observed. Under this interpretation, a crossing establishes that the cohort did not meet the declared target. If $b$ instead describes a superpopulation parameter, a long-run average, or an aspirational goal, a crossing establishes only that the realised cohort exceeded $b$. Inference about the underlying process requires a separate statistical model and lies outside the record-based check.

# 4. Bounds from operational records

**Proposition 1 (reversal bound).** Under A0–A3, $R \le F$.

*Proof.* Each counted reversal identifies a decision in $D$ (A0), distinct from the others (A1), that the platform's authoritative process judged wrong on the original evidence under the same standard (A2, A3). The reversed decisions are therefore a subset of the wrongful ones. ∎

**Proposition 2 (miss bound).** Under the analogues of A0–A3 for reports, $C \le FN$. If a $\delta$ share of counted report-driven corrections are not genuine misses, then $FN \ge (1 - \delta)C$. Therefore, for a maintained upper bound $\bar{\delta}$,

$$FN \ge (1 - \bar{\delta})C.$$

*Proof.* Symmetric to Proposition 1. Each counted correction resolves to a distinct item in the slice that was left up and that the platform's authoritative process subsequently removed under the same evidence, policy version and decision bar, so the corrected items are a subset of the missed violations. The $\delta$ adjustment discounts those corrections that are not genuine concessions. ∎

**Proposition 3 (robustness to selection and to stopping).** Let $R_t$ be the reversals recorded by time $t$. Then $R_t \le R \le F$ for every $t$, whatever process generates appeals, and however that process depends on item characteristics, user characteristics, or on $F$ itself.

*Proof.* $R_t$ is a non-decreasing count of realised events, and Proposition 1 does not reference the mechanism generating them. The inequality holds on every path, so no rule for choosing when to look can inflate it. ∎

Proposition 3 is why these bounds are beneficial, and it is the paper's least obvious claim. The standard objection to user feedback — that it is selective and unrepresentative — is an objection to using it as an *estimator*. **Selection makes a bound loose; it does not reverse it.** Because the inequality holds pathwise rather than in probability, continuous monitoring requires nothing: there is no error rate to expend, so there is nothing to correct for repeated looks.

**A3 is one-sided.** Only reversals that are *not genuine concessions* threaten Proposition 1; incorrectly upheld decisions leave $R$ smaller and the bound merely looser. Writing $\varepsilon$ for the proportion of counted reversals that are not genuine concessions under A2–A3 — whether through adjudication error, evidence supplied after the fact, a policy change, a goodwill restore, or an account-level action —

$$F \geq (1 - \varepsilon)R.$$

§6.2 treats the measurement of $\varepsilon$; §5 shows how much of it the check tolerates.

# 5. The record-based check

**Claim.** Let $\bar{\varepsilon}$ be a maintained upper bound on $\varepsilon$. If $(1 - \bar{\varepsilon})R/D > b$, the realised cohort does not meet the declared precision target under A5.

*Proof.* $F \geq (1 - \bar{\varepsilon})R$ by §4, so $1 - p = F/D \geq (1 - \bar{\varepsilon})R/D > b$. ∎

Given a declared error budget $b$ and assumptions A0–A3 and A5, the two observed counts $D$ and $R$ suffice: one division and one comparison, with no sample of decisions, no reviewers, no interval, and no assumption about who appeals.

*Illustration.* One million removals, 47,000 reversals: 4.7% of removals conceded wrong against a 3% error budget. The finding survives any false-reversal rate below $1 - bD/R$ = **36.2%**. A platform would have to be wrong about more than a third of its own granted appeals for the conclusion to fail.

**On $\bar{\varepsilon}$.** Two cases must not be conflated. Where $\bar{\varepsilon}$ is asserted — a governance parameter, or a conservative constant — it is a **sensitivity assumption** rather than an estimated bound, and the check carries no sampling error. The honest way to report a finding that rests on one is to publish the **break-even value** of $\varepsilon$ at which the finding would fail, so that a reader can judge the assumption directly; on the figures above that value is 36.2%. We reserve the term *upper bound* for a value supported by a one-sided confidence interval estimated from a sample of granted appeals, in which case the check inherits that sample's error rate and this must be declared. §6.2 recommends the asserted case as the default: the break-even value is usually wide enough that a platform can act on a finding without measuring $\varepsilon$ at all.

**Why this is more beneficial than its arithmetic.** The check is record-based: it compares the platform's own operational records with the platform's own published target. It does not depend on audit sampling or on the appealing population being representative.

That is not the same as unrebuttable, and we do not claim it is. The available rebuttals are: the reversal count is wrong; the reversal and decision populations cannot be reconciled (A0, §6.3); the cohort accounting is calendar-based rather than cohort-based (§3); the appeal standard is more permissive than the audit standard (A2); $\varepsilon$ exceeds the tolerance; or the target is aspirational rather than a statement of measured performance. Each is inspectable, and none requires a model of appellant behaviour. What the check removes is the *sampling* dispute, not disputation as such: it trades an argument about whether a sample was representative for an argument about whether two record systems describe a common population. That is a real gain, because the second kind is settled by inspection and yields a reportable defect either way (§6.3), but it is a substitution and not an escape. That is the property we are claiming.

A crossing should first be recorded as a consistency contradiction. Verify population alignment, disposition classification, data lineage and target provenance before attributing it to enforcement quality. If those checks hold, the target-miss conclusion follows. A non-crossing does not validate quality.

**Early detection.** Because a partial count is still a bound (Proposition 3), the check runs continuously and a firing is actionable immediately. Appeal-window maturity is not a precondition for a positive finding; it bears only on interpreting a negative one. The worse the system, the sooner the breach appears: a slice heading for 47,000 reversals against a 3% error budget shows it once about 64% of appeals have arrived; one heading for 150,000 shows it at 20%. In practice, a reported figure can be contradicted after tens of days rather than after the statutory window closes. A six-month window (DSA Art. 20) delays only the all-clear, which §9 argues was never available.

**When an audit becomes necessary.** If $(1 - \varepsilon^{-})R/D \le b$, operational records do not settle the question and the unappealed remainder is unconstrained. Only sampling speaks to it. §7 establishes what sampling can and cannot then contribute.

# 6. What threatens validity, and what does not

| Property | Threatens validity? | Affects how often a finding appears? |
|---|---|---|
| Appeal participation | No (Prop. 3) | Yes |
| Appeal latency | No (Prop. 3) | Yes |
| Second-review error | Yes | Yes |
| Population mismatch (A0) | Yes | Yes |

## 6.1 Participation and latency

Both change the magnitude of $R$; neither changes the direction of the inequality. Low participation and early measurement make findings rarer, not false. State this plainly in any evaluation: the intuition that appeals data is "too biased to use" is correct for estimation and wrong for falsification.

Two consequences. Report `D_a/D`, the share of removals that were even eligible for appeal, alongside `R/D`; it is the measurable handle on how much silence to expect. And report the observed count as a bound, never inflated by a projection; if a projected total is useful operationally, label it an estimate and keep it separate.

## 6.2 Second-review error

Only reversals that are not genuine concessions break Proposition 1, at rate $\varepsilon$. This matters because reversals are compared against a quantity of order $D(1 - p)$: with one million decisions, 99% precision implies about 10,000 wrongful removals and 98% implies about 20,000, so a one-point movement doubles the population being bounded.

Platforms that wish to measure $\varepsilon$ should re-adjudicate a sample of *granted* appeals against the original standard, blind to the appeal outcome. This is a reportable quantity, not a precondition: a platform that has not measured $\varepsilon$ can still act on a finding that survives any plausible value. A pooled measurement, however, bounds only the average rate across the slices it pools; it does not bound any individual slice. Rates may differ by policy area, market, language and surface, so slice-level use requires slice-specific bounds, a justified exchangeability assumption, or a conservative worst-case value, and whichever is relied on must be stated.

**Shared blind spots.** A confidence interval covers sampling variation only. If the automated system, the second review and the audit share a systematic misreading of policy, the audit's $\hat{p}$ is inflated *and* $R$ is deflated simultaneously, so the check becomes unfireable and the instrument fails silently. This is the mechanism most likely to produce $\bar{p} > p$, the condition §7.2 characterises. Evaluator diversity, adversarial test sets and refreshed holdouts reduce but do not eliminate it. If the automated system has been tuned against a known audit set, evaluation leakage produces the same signature.

Shared training and $\varepsilon$ point in opposite directions: shared blind spots may suppress valid reversals, while mistaken or non-merit grants inflate $R$. This paper does not establish which effect dominates.

## 6.3 Population mismatch: verifying A0

The audit's denominator is decisions of a particular kind. Appeal logs are keyed by case and may include appeals against other enforcement paths, other windows, or other deduplication rules. If $R$ is drawn from a superset of $D$, the inequality fails in exactly the direction that manufactures findings.

The requirement is verified one-to-one lineage from each counted reversal to a decision in the audited population, plus an explicit inclusion rule. The concrete failure to look for is a reason-incomplete restore table that mixes user appeals with bulk sweeps, required restores and classifier-rollback restores.

In practice this is the likeliest explanation for a firing check and should be the first hypothesis. It is also a reportable defect in itself: two counts that cannot be reconciled to a common population mean the published precision figure describes a population the platform cannot identify.

# 7. Why the audit-based crossing is a specification check, not a quality test

When an audit is available, the natural check is to scale its estimate to the population, obtain an implied maximum for $F$, and fire when reversals exceed it:

$$(1 - \bar{\varepsilon})R > D(1 - \bar{p}).$$

This is the check a practitioner builds, and it is the one we built first. It does not work.

## 7.1 The result

**Proposition 4 (crossing-bound firing is bounded by miscoverage).** Let $L$ be an observed lower bound on an unknown quantity $Q$, with $L \le Q$ almost surely. Let $U$ be an upper confidence bound for $Q$ with one-sided coverage $1 - \alpha$ for the realised value of $Q$. Then for every value of $Q$,

$$P(L > U) \le \alpha.$$

*Proof.* $\{L > U\} \subseteq \{Q > U\}$ pathwise, because $L \le Q$. The right-hand event is failure of coverage, which has probability at most $\alpha$. ∎

The argument is elementary, holds on every path, and is therefore indifferent to which of $L$ and $U$ is random and to any dependence between them. It is not new: this is the crossing-bound specification test of the partial-identification literature (Bugni, Canay and Shi 2015), and we state it only to apply it. It would be helpful to make three conditions explicit because applications routinely violate them. $L \le Q$ must hold almost surely, not on average — which is why the application below uses $L = (1 - \bar{\varepsilon})R$ rather than $R$. Coverage must be one-sided, frequentist and pointwise in $Q$; Wald-type intervals for precision undercover near $p \approx 0.99$ at moderate $n$, exactly this regime (Cai 2005, fig. 1), and under asymptotic-only coverage the conclusion is asymptotic only. And $U$ need not come from the same data as $L$ — only from the same estimand.

**Corollary 1 (uniformly limited firing probability).** Under the maintained assumptions, the crossing rule has firing probability at most $\alpha$ for every value of $Q$. It therefore cannot have power exceeding $\alpha$ against any quality state represented only by a change in $Q$. A firing indicates that coverage failed or that at least one maintained bound or assumption failed; it is not, by itself, evidence that moderation quality is low.

The result establishes a uniform upper bound on firing probability. It does not establish equality of firing probabilities across values of $Q$, nor formal ancillarity in the statistical sense.

We avoid the slogan "power equals size". The result gives rejection probability *at most* $\alpha$, not equal to it; because $\{L > U\}$ is typically a strict and much smaller subset of $\{Q > U\}$, the realised firing rate is usually far below $\alpha$ and can be zero.

## 7.2 What it does test

Firing implies that coverage failed or that A0–A3 failed. The procedure is therefore **a conformance test on the measurement and its assumptions**, not on moderation quality, and a platform whose quality is deteriorating while its audit correctly tracks the deterioration will see the check pass indefinitely, because the threshold $D(1 - \bar{p})$ widens in step.

A firing is consequently not a "false alarm" — §6.3 gives a case where it is a real and reportable defect — but it is uninformative about quality. Both facts are beneficial to know; they are not the same; and a system reported as healthy by this check has not been shown healthy in any sense.

**Scope.** Nothing in the proof is specific to moderation. Any procedure comparing an observed lower bound against a confidence-based upper bound for the same quantity inherits the result. Regulatory regimes that require both an accuracy figure and a complaint-outcome figure invite exactly this comparison, which is why we state it generally. This diagnosis, rather than the inequality, is the contribution.

**The conservatism is the culprit, not the comparison.** Comparing $(1 - \bar{\varepsilon})R$ against the point estimate $D(1 - \hat{p})$ is *not* covered by Proposition 4 and does retain power, at roughly 50% size at the boundary. What destroys the audit-based check is the one-sidedness of $U$, not the act of comparing two quantities.

## 7.3 Historical references and drift

Carrying a confidence bound forward from an earlier period changes the question from contemporaneous conformance to temporal drift. Such a comparison may acquire power against deterioration, but the earlier bound is no longer an upper confidence bound for the current quantity.

Its false-firing probability depends on uncertainty in both periods, stability of the decision population, changes in reversal capture, and the adjustment for reversal error. A calibrated drift procedure therefore requires a separate null hypothesis and an explicit two-period or sequential design. We leave that problem to future work.

## 7.4 Statistical status and multiplicity

$\bar{p}$ is a confidence bound, not a population maximum. A firing is a level-$\alpha$ rejection of the conjunction of A0–A4 and the audit's interval, not a proof of impossibility, and it does not identify which conjunct failed. Descriptions of such a result as proving a reported figure cannot be true overstate it.

Run per slice and per period, the testing family must be declared. If the per-slice firing rate were at its ceiling of 2.5% one-sided and slices were independent, the probability of at least one false alarm would be 39.7% over 20 slices and 92.0% over 100. Proposition 4 gives *at most* that rate, and §7.1 notes the realised rate is usually far lower, so these figures bound the exposure rather than describe it. The check in §5 is immune to this: with an asserted $\bar{\varepsilon}$ it has no sampling error and therefore no sampling false-alarm rate, though its assumptions remain the object of scrutiny.

# 8. A recall ceiling with no participation floor

Recall cannot be measured without finding violations among items left up. Confirmed reports establish a minimum miss count and therefore a maximum recall. With $T$ correct removals and $FN \geq (1 - \bar{\delta})C$ (Proposition 2),

$$\text{Recall} \leq T / (T + (1 - \bar{\delta})C).$$

Under A3 without relaxation, $\bar{\delta} = 0$, and this reduces to $T/(T+C)$.

For this to be a valid ceiling, $D$ must capture all removals of violating content in the slice. Violations removed by an enforcement path outside $D$ are true positives excluded from $T$, which makes the computed ceiling too tight and the claim too strong.

**This requirement is stronger than its appeals-side analogue, and is the recall result's principal fragility.** A0 for appeals asks that reversals resolve *into* $D$; the condition here asks that $D$ exhaust every enforcement path that removes violating content in the slice, while $C$ is by construction a count of items $D$ did not act on. Platforms typically operate several such paths — proactive classifiers, integrity sweeps, account-level actions, required removals, and surface-specific pipelines — and a slice defined by one of them will exclude true positives removed by another. The verification is therefore an enumeration of enforcement paths and a demonstration that the excluded ones remove nothing in scope, not merely a lineage join. A ceiling computed without it is not conservative; it is invalid in the direction that manufactures findings.

## 8.1 Bounding T correctly

The ceiling increases in $T$, so a valid ceiling requires a valid **upper** bound on $T$. The audit supplies the one-sided confidence bound $T \leq D \cdot p^{+}$. The operational record supplies

$$T \leq D - (1 - \bar{\varepsilon})R,$$

because at least $(1 - \bar{\varepsilon})R$ counted reversals are genuine concessions under the relaxed model. Combining the two gives

$$U_T = \min(D \cdot p^{+}, D - (1 - \bar{\varepsilon})R).$$

When A3 is maintained without relaxation, $\bar{\varepsilon} = 0$ and the operational term reduces to $D - R$. The audit component is a confidence bound at the declared coverage level; the operational component is deterministic conditional on the maintained upper bound for $\varepsilon$.

Substituting $T = \hat{p} \cdot D$ is the natural move and is wrong: exact for true precision, merely an estimate for sampled precision, and the resulting quantity is not a bound and can be violated.

Counterexample: $D = 100$, all removals correct, $C = 5$, $\hat{p} = 0.95$ gives a "ceiling" of 95% while true recall is 100/105 = 95.24%.

*Illustration.* Taking $\varepsilon^- = 0$ and $\delta^- = 0$, let $D = 1{,}000{,}000$, $C = 140{,}000$, $p^+ = 98.76\%$ and $R = 47{,}000$: the audit path gives 87.58%, the operational path 87.19%, and discarding the audit entirely ($T \le D$) still gives 87.72%.

A 95% target is unreachable under all three, the last using no audit at all.

## 8.2 Why this survives where §7 does not

The recall ceiling has no participation floor. Every confirmed miss tightens it monotonically, whatever the reporting rate. Low participation here makes the ceiling looser rather than uninformative about quality in the way Corollary 1 describes for the crossing rule. It does, however, have a classification-error analogue to $\varepsilon$. A later removal is an authoritative platform disposition, but its direction alone does not establish that the original decision was wrong. This uncertainty is represented by $\delta$, and the recall ceiling uses the adjusted lower bound $(1 - \delta^-)C$.

Scale-free form: writing $r$ for the recall target, the target is breached whenever $(1 - \delta^-)C > U_T(1 - r)/r$. When the audit upper bound $U_T = D \cdot p^+$ is used, this becomes $(1 - \delta^-)C/D > p^+(1 - r)/r$. For a 95% target, precision near 1 and $\delta^- = 0$, that means confirmed misses above about 5.3% of removals: a single number to watch.

**A drift mode with no adversary.** The ratio $C/D$ is normalised by removals, not by the violating population. Anything that shrinks $D$ for reasons unrelated to quality — a narrowed policy scope, a decline in prevalence, or a surface migrating to an enforcement path outside the slice — raises $C/D$ and can fire the check while enforcement is unchanged. This is not gaming, and §9.2 does not cover it. Any evaluation should therefore report $D$ as a level alongside the ratio, and treat a firing accompanied by a falling $D$ as a scope change to be excluded before it is read as a quality finding.

# 9. Future Evaluation

## 9.1 What a clean result does not mean

A finding is informative. Its absence is not. Lower participation, a shorter window, a smaller sample and a wider interval all make findings rarer without making the system better. A slice producing no finding is **unexamined, not clean**, and should be reported as "no contradiction detected."

Evaluations should distinguish three outcomes, not two: **finding**, **no finding**, and **uninformative**. Reporting the third as if it were the second is the failure this paper most wants to prevent, and the distinction is only meaningful if the boundary is **declared before the data are extracted**. We therefore treat diagnostic reach as an element of the declared testing family (§7.4, §12): the rule is fixed once, in advance, and evaluated separately for every slice with that slice's own counts.

Each check is uninformative for a different reason, and each has an observable criterion.

- **§5, precision.** Only appeal-eligible decisions can be reversed, so $R \le D_a$ and the largest value the check's left-hand side can take is $(1 - \bar{\varepsilon}) \cdot D_a / D$. Where that quantity is no greater than $b$, the check cannot fire however poor enforcement is, and the slice is uninformative as a matter of arithmetic rather than of judgment. Both inputs are already required by §6.1.
- **§8, recall.** Nothing caps $C$ as $D_a$ caps $R$, which is the other face of having no participation floor: the ceiling tightens at any reporting rate, so the recall result is never unreachable on reach grounds. Its criterion is instead **maturity** — the arrival fraction of the report window, which §6.1 also already requires.
- **§7, the audit crossing.** Uninformative when the interval is wide enough that the implied threshold cannot be crossed by any attainable reversal count.

We do not prescribe the threshold at which a slice becomes uninformative. §11.4 argues that a reporting regime should compel disclosure of a method rather than mandate one, and fixing a figure here would contradict that. What we require is that a threshold be declared in advance and published with the result. Where one threshold is applied across slices that differ in policy area, market, language or surface, it needs the same justification §6.2 requires of a pooled $\bar{\varepsilon}$: slice-specific values, an argued exchangeability assumption, or a conservative worst case.

**This does not reintroduce a participation model.** The *validity* of the bounds requires none, which is Proposition 3. Diagnostic *reach* is about participation and always was — §6.1 already states that participation governs how often a finding appears. Declaring a reach threshold quantifies something the paper already concedes matters; it adds no assumption on which any bound depends.

It would be helpful to make one consequence explicit. A generously set $b$ raises the smallest breach the check could have surfaced, which pushes slices into the uninformative class. A platform cannot therefore achieve quiet passes by declaring a lax target: doing so visibly shrinks the share of its own estate that the instrument can examine. This is the partial answer to the gaming hazard of §11.4.

The asymmetry is not a defect to engineer away. It follows from the bounds being one-sided, the same property that makes them robust to unknown selection. A framework that could also certify health would need exactly the participation model Propositions 1–3 avoid.

## 9.2 Incentives

Every input governing whether the §5 and §7 checks fire — sample size, confidence level, slice boundaries, deduplication rules, maturity window, and the standard applied when granting appeals — is controlled by the organisation being measured, and the gradient runs the wrong way: a smaller audit widens the interval and suppresses findings. These parameters must be fixed externally and in advance.

One apparent remedy is not one: granting fewer appeals reduces $R$ and silences both §5 and §7 without correcting a single wrongful removal. A falling reversal count requires explanation and is never by itself an improvement. The same applies to a falling confirmed-report count.

A related hazard is definitional. Where a platform is free to choose how it operationalises its own accuracy indicator, it may choose a definition derived from appeal outcomes. The declared target is then not independent of the count meant to test it, and the check in §5 becomes tautological rather than merely quiet. Any evaluation should confirm that $b$ is set independently of $R$. §11 shows that the regime does not require this, so the condition must be established by the evaluation itself rather than assumed from compliance.

### 9.3 Minimum analysis record

Any reported result should carry enough information for another analyst to reconstruct it without undocumented operational knowledge:

- The complete slice definition and cohort dates, and confirmation of cohort rather than calendar accounting; the appeal and report maturity cutoff;
- $D$, $D_a$, $R$, $C$, $b$, and where an audit was used, $\hat{p}$, $[p^-, p^+]$, $n$, and $\alpha$;
- Model and policy versions at the automated decision and at second review;
- Sampling, weighting, denominator, deduplication, and case-construction rules;
- The lineage-verification result for A0;
- The values of $\varepsilon^-$ and $\delta^-$, whether each was asserted or estimated, and if estimated, the corresponding sample, method and confidence interval;
- The arrival fraction behind any projected figure;
- The declared diagnostic-reach threshold and this slice's computed reach (§9.1);
- The declared testing family and multiplicity correction;
- The resulting classification: finding, no finding, or uninformative.

## 10. Limitations and exclusions

**The framework cannot certify health.** It is a falsifier. §9.1 states the consequence: absence of a finding is not evidence of quality.

**It cannot evaluate discrimination or discriminatory harm**, and must not be presented as evidence of fairness, non-discrimination, civil-rights compliance, disparate impact, or equal opportunity in housing, employment, credit or any other domain. Those require comparisons across people, exposure and outcomes that these counts do not support.

**It does not establish true recall, total harm, or system safety.** A rejected appeal or report does not establish bad faith. Results are specific to a declared slice and must not be pooled across harm types before the checks are run separately.

**Unequal access to recourse.** Participation cannot make a finding false, but it does mean the instrument's protection is distributed unevenly: populations that appeal less are less protected by it. This is measurable from the `D_a/D` and arrival statistics §6.1 already requires, and should be reported rather than left implicit.

**We report no results.** §11 analyses the EU's oversight landscape and regime-level aggregates to establish that the framework's independence precondition is not currently required of anyone; it measures no provider's moderation quality and reports no provider-specific finding. Proposition 4 and Propositions 1–3 are not empirical claims. The practical claims are that appeals can be reconciled to an audited population (A0), that `ε` can be bounded, and that platforms have a declared `b` set independently of `R`. A single retrospective slice would test all three, and §12 registers exactly how. We state this as the boundary of the present contribution rather than as a deficiency in it: the propositions, the negative result, the specification and the conformance profile are established analytically and do not await data.

The framework would be falsified as *useful*, though not as *incorrect*, if reversal rates in practice sit far below error budgets in every slice — in which case §5 never fires and what remains is §7's conformance check and §8's ceiling.

# 11. What the reporting regime makes falsifiable

Sections 5 and 8 require a declared target `b` set independently of the operational counts (§9.2). Whether such a target exists anywhere is not a question about any individual provider's conduct; it is a question about what the reporting regime obliges providers to publish, and it can be answered from the oversight landscape alone. We examine the European Union's transparency regime because it is the most developed, and because its structure is likely to be copied.

Article 15(1)(e) DSA requires a provider to report "any use made of automated means for the purpose of content moderation, including a qualitative description, a specification of the precise purposes, indicators of the accuracy and the possible rate of error of the automated means used in fulfilling those purposes, and any safeguards applied". Article 42(2)(c) requires the accuracy indicators broken down by official language. Commission Implementing Regulation (EU) 2024/2835 harmonises the templates and specifies the field further: it must describe the indicators "namely the accuracy, precision and recall for automatically removed content", and must include "a description of the input criteria that produce these indicators and the methodology used to calculate such indicators".

Two observations follow, and they point in opposite directions.

## 11.1 What the regime now does well

The harmonised template names *precision and recall* explicitly, which are the two quantities bounded in §5 and §8. It also compels disclosure of the methodology by which each indicator was produced. That second requirement is, in substance, the provenance rule recommended in §12: a reader can now be told whether a published figure came from a sample, from a complaint record, or from a qualitative assertion. We note this because the framework does not need to argue for it—the regime has adopted it—and because the remaining gap is sharper once it is set aside.

## 11.2 Accuracy is a term the regime does not define

Neither the Article nor the template fixes what "accuracy" is to be computed from. Sekwenz, Wagner and de Bruijn (2026) likewise document non-equivalent labels for automated-accuracy measures in platform reporting, including "automation overturn rate", "specific accuracy" and "error rate". Three definitions are admissible under the text as written, and they are not equivalent:

| Basis for the indicator | Resource requirements | Consequence for falsifiability |
|---|---|---|
| Complaint outcomes: accuracy as the share of decisions not overturned | Near zero; the records already exist | The indicator becomes a function of complaint participation, eligibility and granting standards. It cannot be contradicted by the complaint record, because it is the complaint record. Granting fewer complaints improves it. |
| Sampled re-review: precision estimated from a sample of actioned content | Material and recurring | Independent of the complaint record, so the bounds of §4 can test it. Carries sampling error, and does not reach low-volume slices. |
| Qualitative description only | None | Nothing to test. |

The first basis is the one the text most readily permits and the simplest to satisfy. Where it is chosen, the check of §5 is not merely quiet but **tautological**: the declared quantity and the evidence against it are the same measurement. This is the failure mode of §9.2, and it is a predictable consequence of leaving the term undefined rather than a defect of any particular filing.

## 11.3 Disclosure is not commitment

All three bases produce a *realised* indicator. None of them, and nothing elsewhere in the regime, requires a provider to declare a precision or recall **target in advance**. The obligation is to disclose what happened, not to commit to what should happen.

This is the gap that matters for everything above. A realised figure cannot be missed, because nothing was promised; there is no `b` for `(1 − ε)R/D` to exceed.

The machinery of §5 is well defined and operationally simple, and on the current regime it has nothing to bind to. Falsifiability was not removed by any provider; it was never required.

## 11.4 What would close it, and what that would require

The minimal addition is a declared error budget: a precision and recall target, for a defined slice, fixed in advance, set independently of the complaint record, and published alongside the realised indicator and the conceded-error counts `R` and `C`. With those elements present, §5 and §8 run on arithmetic and require no new instrumentation.

We set out the objections rather than dismiss them, because a specification that ignores its own downsides will not be adopted.

- **A declared target creates exposure for missing it.** The rational response is to declare a target low enough to be safe. A target set generously is unfalsifiable by generosity: the check never fires and nothing is gained. The mitigation is to require that the published target be the figure used internally for goal-setting, and that any revision be published with a reason.
- **Per-slice targets multiply reporting burden.** Policy area by market by language by surface is a large cross-product, and the requirements for disclosure can exceed the requirements for measurement. The mitigation is to require targets only where volume makes the check informative, and to report coverage—the share of slices carrying a target—as a figure in its own right.
- **Prescribing one methodology would freeze measurement.** A mandated method that is wrong is worse than plural methods whose provenance is disclosed. This supports the regime's existing choice to compel disclosure of the method rather than the method itself. The target requirement is separable from the methodology requirement and should remain so.
- **Sampled precision does not scale down.** Any target regime resting on it leaves small markets, new surfaces and low-volume policy areas unexamined, which is where reporting is weakest already. The recall bound of §8 is the partial answer, since a confirmed miss tightens the ceiling at any reporting rate.
- **A declared target invites definitional drift in the slice.** A provider that defines the population to which the target applies can move the population instead of the target. The slice definition must therefore be fixed with the target and published with it.

## 11.5 A conformance profile

The framework is useful only if a third party can tell whether a published account is **checkable**. We state the minimum as a profile, so that conformance is a property of the disclosure rather than a matter of judgment. A published account is **checkable** for a slice when it contains:

1. the slice definition—policy area, market, language, surface, enforcement path and cohort window—with confirmation of cohort rather than calendar accounting;

2. D, the decisions in the slice, and D_a, those eligible for complaint;
3. R, conceded reversals of decisions within that cohort, deduplicated, and C, confirmed misses;
4. b, declared in advance, with a statement that it was set independently of R, the date it was fixed, and **which interpretation it carries** — a target for the realised cohort, or a long-run or aspirational figure (A5). Without that statement a reader cannot tell whether a crossing is conclusive;
5. the provenance of every accuracy indicator: sample, complaint record, or assertion;
6. the maintained values for ε and δ, whether each is an asserted sensitivity assumption or an estimated bound, and the break-even value of each where a finding is reported;
7. the declared diagnostic-reach threshold and the slice's computed reach;
8. the resulting classification: finding, no finding, or uninformative.

Items 1 to 3 and item 5 are already obtainable, or nearly so, under the harmonised template. Item 4 is the substantive addition. Items 6 to 8 are matters of reporting discipline rather than new measurement. Nothing in the profile requires a provider to disclose a quantity it does not already compute, which is the property that makes adoption plausible.

## 11.6 The records are large enough to bind

It is beneficial to propose a specification only if the underlying counts are substantial. Reported aggregate figures for the regime indicate that they are: reviewing two years of application, the European Commission reported that roughly 165 million content-moderation decisions were contested through providers' internal complaint mechanisms, of which about 30 percent were overturned—approaching 50 million conceded reversals in aggregate (European Commission 2026). These are regime-level aggregates across providers and cannot be used to check any individual target, which is the purpose of §3's prohibition on pooling across slices. They do establish that R is not a marginal quantity.

**An external instrument for ε.** §6.2 treats ε as requiring a provider to re-adjudicate its own granted complaints. The Article 21 out-of-court dispute settlement mechanism is the first body with standing to review such decisions from outside the provider's measurement apparatus, and the European Commission reported that these bodies overturned provider decisions in 52 percent of closed cases among more than 1,800 disputes reviewed in the first half of 2025 (European Commission 2026). The direction matters: that is evidence about decisions a provider *upheld* and should not have, which leaves R an understatement and Proposition 1 intact. It does not bound ε, which concerns grants that should have been refusals, and the two must not be conflated. Caseloads remain small relative to complaint volumes and the referring population is self-selected. We record the mechanism as the most promising external instrument available, not as a measurement.

## 11.7 Scope of this section

This section analyses the EU's oversight landscape and regime-level aggregates. It makes no claim about the accuracy, the enforcement quality or the compliance of any named provider, and reports no provider-specific finding. The unit of analysis is the specification. Where we observe that a definitional choice is available and not operationally burdensome, that is a statement about what the text permits, not an allegation that any provider has chosen it improperly or that any published figure is wrong.

# 12. Evaluation protocol

This section specifies a protocol for a future empirical evaluation. The present paper develops the theoretical framework and analytical bounds; it does not execute the protocol, analyze operational records, or report empirical results. The protocol is stated in advance so that any future evaluation can define its sample, variables, and decision rules before operational data are examined, reducing the scope for post-hoc adjustment. Execution of the protocol and any resulting findings are outside the scope of this paper.

1. **Declare** the slice, the unit of analysis, the automated decision rule, $b$ and its interpretation under A5, the maturity window, $\varepsilon^{-}$, $\delta^{-}$, $\alpha$, the diagnostic-reach threshold (§9.1), and the testing family, before extracting data.
2. **Verify A0** by resolving every counted reversal to a decision in $D$; report the resolution rate and the disposition of unresolved records. Confirm cohort rather than calendar accounting.
3. **Confirm $b$ is independent of $R$** (§9.2).
4. **Bound $\varepsilon$ and $\delta$**, either through explicitly stated sensitivity assumptions or through blinded re-adjudication samples against the original standard. Pre-specify the sampling design and the one-sided confidence-interval method. Declare which route was used for each.
5. **Run §5** on the operational counts. Record finding / no finding / uninformative.
6. **Where §5 does not settle it**, run the audit and the §7 comparison, reporting it explicitly as a conformance check on the measurement.
7. **Compute the §8 ceiling** using the adjusted upper bound on $T$ and the adjusted lower bound on $FN$.
8. **Repeat prospectively** on a later cohort without changing registered rules.
9. **Record the provenance of $b$** in the published account, stating whether it is a target declared in advance or an indicator computed from outcomes, and which interpretation of A5 applies, so that a reader can tell whether the check was capable of firing at all (§11).

Primary outcomes: A0 resolution rate; the values and provenance of $\varepsilon^{-}$ and $\delta^{-}$; whether §5 fires and at what fraction of appeal maturity; interval calibration; and stability across harm-specific cohorts.

**Negative and positive controls.** Simulate or replay with known $F$, $\varepsilon$ and participation; inject audit bias, join errors, duplicated appeals and second-review error as the alternatives §7 is actually able to detect; and include a pure precision regression measured correctly by both audit and appeals, to demonstrate that §7 correctly stays quiet while a separate target test fires.

# 13. Conclusion

A platform's own records constrain the quality figure it publishes, in one direction, without any sample. Reversals bound wrongful removals from below and confirmed reports bound misses from below, each discounted by the share of counted concessions that may not be genuine, and either can contradict a target declared in advance for the realised cohort on its own. Because these are bounds rather than estimates, the objection that routinely disqualifies user feedback — that the people who complain are not representative — does not apply: selection makes the bound loose, never wrong.

The same asymmetry limits what the approach can deliver. It can show that a target was not met; it cannot show that one was. The natural attempt to close that gap by bringing in the audit does not work, for a reason that is elementary once stated and that applies well beyond moderation: an alarm that compares an observed lower bound against a confidence-based upper bound for the same quantity tests the measurement, not the system.

What remains is narrow and not operationally burdensome, and we think it is the right trade. Two integers a platform already holds can establish that a published target was missed without relying on audit sampling. A finding triggers a bounded investigation; its absence does not validate quality, and where the records could never have reached the declared error budget the slice must be reported as uninformative rather than clean. Whether these bounds are informative in practice is an empirical question beyond the scope of this paper; §12 specifies a prospective protocol for future evaluation.

# Author contributions

Aayush Agarwal: Conceptualization, methodology, formal analysis, writing—original draft, and project administration.

Daron Veer Singh Sohi: Methodology, formal analysis, and writing—review and editing.

Linda Cornelison: Conceptualization, governance and domain framing, and writing—review and editing.

# References

1. Ames, D., Handan-Nader, C., Ho, D. E., and Marcus, D. (2019). Quality Review of Mass Adjudication: A Randomized Natural Experiment at the Board of Veterans Appeals, 2003–16. *The Journal of Law, Economics, and Organization*, 35(2), 239–288.
2. Begg, C. B., and Greenes, R. A. (1983). Assessment of diagnostic tests when disease verification is subject to selection bias. *Biometrics*, 39(1), 207–215.
3. Bugni, F. A., Canay, I. A., and Shi, X. (2015). Specification tests for partially identified models defined by moment inequalities. *Journal of Econometrics*, 185(1), 259–282.
4. Cai, T. T. (2005). One-sided confidence intervals in discrete distributions. *Journal of Statistical Planning and Inference*, 131, 63–88. https://doi.org/10.1016/j.jspi.2004.01.005
5. Chandra Sekar, C., and Deming, W. E. (1949). On a method of estimating birth and death rates and the extent of registration. *Journal of the American Statistical Association*, 44(245), 101–115.
6. Chernozhukov, V., Lee, S., and Rosen, A. M. (2013). Intersection bounds: estimation and inference. *Econometrica*, 81(2), 667–737.
7. European Commission (2024). Commission Implementing Regulation (EU) 2024/2835 of 4 November 2024 laying down templates concerning the transparency reporting obligations of providers of intermediary services and of providers of online platforms under Regulation (EU) 2022/2065.
8. European Commission (2026). *Two years of Digital Services Act allows 50 million content moderation decisions by platforms to be reversed*, 20 February 2026. https://digital-strategy.ec.europa.eu/en/news/two-years-digital-services-act-allows-50-million-content-moderation-decisions-platforms-be-reversed
9. European Union (2022). Regulation (EU) 2022/2065 on a Single Market for Digital Services (Digital Services Act), in particular Articles 15, 20, 21 and 42.
10. Fienberg, S. E. (1972). The multiple-recapture census for closed populations and incomplete contingency tables. *Biometrika*, 59(3), 591–603.
11. Horvitz, D. G., and Thompson, D. J. (1952). A generalization of sampling without replacement from a finite universe. *Journal of the American Statistical Association*, 47(260), 663–685.
12. Imbens, G. W., and Manski, C. F. (2004). Confidence intervals for partially identified parameters. *Econometrica*, 72(6), 1845–1857.
13. International Auditing and Assurance Standards Board. ISA 530, *Audit Sampling*.
14. Lakkaraju, H., Kleinberg, J., Leskovec, J., Ludwig, J., and Mullainathan, S. (2017). The selective labels problem: evaluating algorithmic predictions in the presence of unobservables. *Proceedings of the 23rd ACM SIGKDD International Conference on Knowledge Discovery and Data Mining*, 275–284.
15. Manski, C. F. (1989). Anatomy of the selection problem. *Journal of Human Resources*, 24(3), 343–360.
16. Manski, C. F. (2003). *Partial Identification of Probability Distributions*. Springer.
17. Pingen, A. (2026). Two Years of the Digital Services Act. *eucrim*, 21(1). ISSN 1862-6947.
18. Public Company Accounting Oversight Board. AS 2315, *Audit Sampling*.

19. Schwemer, S. F. (2024). Decision Quality and Errors in Content Moderation. *IIC – International Review of Intellectual Property and Competition Law*. https://doi.org/10.1007/s40319-023-01418-4
20. Sekwenz, M.-T., Wagner, B., and de Bruijn, H. (2026). From Reports to Reality: Testing Consistency in Instagram's Digital Services Act Compliance Data. *ACM Journal on Responsible Computing*, 3(3), 1–27. https://doi.org/10.1145/3821526
21. Wei, J. T.-Z., Zufall, F., and Jia, R. (2024). Operationalizing content moderation "accuracy" in the Digital Services Act. *Proceedings of the AAAI/ACM Conference on AI, Ethics, and Society (AIES)*. arXiv:2305.09601.